# Patient-specific AI for generation of 3D dosimetry imaging from two 2D-planar measurements.

Alejandro Lopez-Montes, Robert Seifert, Astrid Delker, Guido Boening, Jiahui Wang, Christoph Clement, Ali Afshar-Oromieh, Axel Rominger, Kuangyu Shi

*Abstract*—In this work we explored the use of patient specific reinforced learning to generate 3D activity maps from two 2D planar images (anterior and posterior). The solution of this problem remains unachievable using conventional methodologies and is of particular interest for dosimetry in nuclear medicine where approaches for post-therapy distribution of radiopharmaceuticals such as 177Lu-PSMA are typically done via either expensive and long 3D SPECT acquisitions or fast, yet only 2D, planar scintigraphy. Being able to generate 3D activity maps from planar scintigraphy opens the gate for new dosimetry applications removing the need for SPECT and facilitating multi-time point dosimetry studies. Our solution comprises the generation of a patient specific dataset with possible 3D uptake maps of the radiopharmaceuticals withing the anatomy of the individual followed by an AI approach (we explored both the use of 3DUnet and diffusion models) able to generate 3D activity maps from 2D planar images. We have validated our method both in simulation and real planar acquisitions. We observed enhanced results using patient specific reinforcement learning (~20% reduction on MAE and ~5% increase in SSIM) and better organ delineation and patient anatomy especially when combining diffusion models with patient specific training yielding a SSIM=0.89 compared to the ground truth for simulations and 0.73 when compared to a SPECT acquisition performed half an hour after the planar. We believe that our methodology can set a change of paradigm for nuclear medicine dosimetry allowing for 3D quantification using only planar scintigraphy without the need of expensive and time-consuming SPECT leveraging the pre-therapy information of the patients.

## I. Introduction

Theranostics has experienced an increased interest during the last years [1]. The additional gamma rays emitted by radionuclides such as [177]Lu in PSMA for prostate cancer treatment can be used to image the biodistribution of the drug. 2D-planar imaging (scintigraphy), is used for monitorization [1], [2] and typically consists of 2 projections (anterior and posterior) obtained with a gamma camera. However, its 2D nature dampers its usability as a quantitative method [1]. SPECT provides a 3D alternative for uptake maps of the radionuclides [2]. However, it is generally a costly and slow procedure, needing multiple bed positions to acquire full body images of the patient and acquisitions at different angles leading to long acquisitions (>1h for full body SPECT) [2]. Some works have explored the use of fast SPECT protocols reducing either the time per view or the number of views (sparse SPECT) [2]. With all, SPECT still results in increased costs for the hospitals and a dedication of resources (expensive scanners) and personnel (e.g. working hours for technicians). Most patients undergoing [177]Lu-PSMA therapy count with a pre-therapy PET/CT using [18]F-PSMA or [68]Ga-PSMA [3]. In this work we leveraged pretherapy (PET/CT) to achieve AI enabled 3D image reconstruction from planar scintigraphy with only 2 (anterior and posterior) projections, which is in principle impractical using conventional (AI and non-AI) methods.

## II. Materials and Methods

### A. Forward 3D-2D problem.

Knowing the 3D distribution of activity ($A_{3D}$) in a patient and the attenuation of x-rays inside the body ($\mu$) the 2D scintigraphy projection ($A_{sc}$) can be obtained from the line integrals [4]:

$$A_{sc}(u,v) = \int A_{3D}(x,y,z) \cdot e^{-\int_0^S \mu(x',y',z')ds'} ds + r(u,v) \quad (1)$$

Where $s$ is line connecting every voxel of the volume $(x,y,z)$ with every pixel $(u,v)$. $r(u,v)$ includes effects such as scattered photons and noise. In this work, we used the common approach in SPECT of approximating equation (1) to a combination of a parallel projection with a PSF (gaussian) blurring given by the collimator properties and distance from the object to the detector [4]:

$$A_{sc}(u,v) = A_{sc}(x,z) \approx$$
$$\int G(x,y,z) * A_{3D}(x,y,z) \cdot e^{-\int_0^y \mu(x',y',z')dy'} dy + r(x,z) \quad (2)$$

We do not consider scattered photons in this work. We modeled $r(x,z)$ as an additive Poisson noise obtained from the distribution $P(\lambda = \bar{A}_{sc}(x,z))$.

### B. Inverse 2D-3D problem with patient priors

The inverse problem for equation (2) is unachievable with only two planar (anterior and posterior) images, however using previous information about the anatomy of the patient and expected uptake, we can approximate the 3D activity maps subject to the 2D observations using AI. Firstly, we used the pre-treatment PET of the patient to generate random realizations of possible uptakes (patient specific). For that, we used the CT-free PET based organ segmentation described in [5]. We generated new distributions for every patient with variations in the uptake of every organ (sampled from a uniform distribution from half to double the original activity), and random rotations ($\pm 5^o$) and translations ($\pm 20$cm). After generating the patient specific dataset, we used AI to recover 3D activity maps from planar inputs. We explored two alternatives (Fig. 1). One is based on supervised learning via Unet and the second approach consisted of an unsupervised generative model based on diffusion models [6]. We compared both approaches with the same networks trained on generic population instead of patient friendly datasets.

### C. Training, sampling and validation of the method

Generic networks were trained on a dataset of 52 total body [18]F-PSMA patients. For patient specific reinforcement we used 100 variations for the patient specific uptake. For the supervised learning we simulated the planar acquisitions using (2) and trained a 3DResUnet architecture described in [7] being the inputs the planar images and the outputs the 3D activity maps.

For the diffusion approach, we trained a score-based generative model [6] with 1000-time steps to generate axial slices representing possible uptakes of PSMA from the patient specific dataset. Then, for the sampling, at every diffusion step we built 3D volumes via slice stacking and we enforced the generation of solutions consistent with the measured projections via Diffusion MBIR [8], [9]. That was achieved by projecting the temporal solutions using (2) with the pre-therapy CT for the attenuation model and comparing them with the actual measured planar images. For all cases, 3D volumes were 128x128x175 and planar images 128x175x2. Both pixel and voxel sizes were 5.7 mm. We validated the methods on simulations (4 cases) obtained from post-therapy $^{18}$F-PSMA PET/CT scans and we tested their performance on a real planar scintigraphy case from $^{177}$Lu-PSMA planar scans.

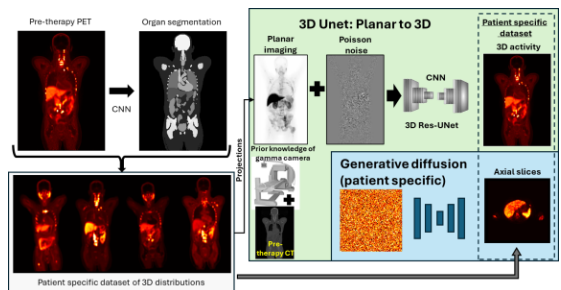

**Fig.1.** Schematic representation of the proposed methodology for training based on patient specific datasets to generate activity maps via supervised learning (3DUnet planar to 3D) or via unsupervised generative diffusion models. In both cases, the network learns the anatomy and expected uptake of the patient

### III. RESULTS

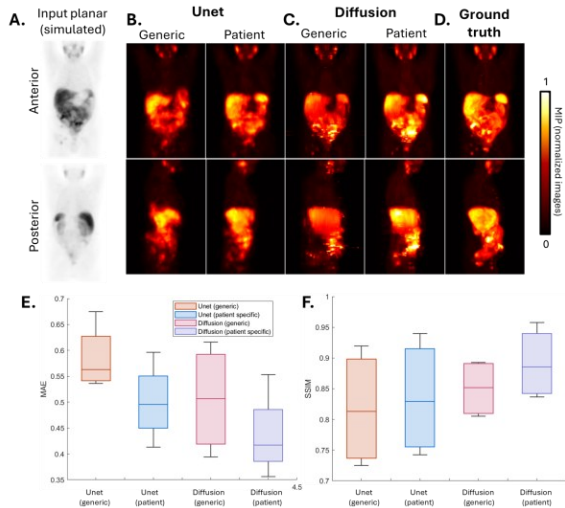

**Fig. 2. A.** Simulated planar inputs from ground truth. **B-C.** Maximum intensity projections (MIP) at 0 (top) and 90 (bottom) degrees for the reconstructed images using the studied methods: Unet (**B**) and Diffusion MBIR (**C**). Generic population dataset training is compared with our patient-specific reinforcement learning for both cases. (**D**) Ground truth 3D activity (planar images in A were simulated from **D**) (**E-F**) Quantification using boxcharts indicating median and interquartile ranges across the cases (**E**). Mean absolute errors across the entire images for the 4 cases used for simulation studies. (**F**). SSIM in the image for the different methodologies.

Fig 2 shows the results of the methods studied on simulations where the ground truth 3D activity from which the planar images were generated is known. Quantifications in (E-F) showed enhanced performance using reinforcement learning based on patient-specific data when compared to the same architectures using generic population training. Average values for MAE were improved from 0.58 to 0.5 for the case of Unet and from 0.51 to 0.43 for the case of diffusion. SSIM (average) improved from 0.81 to 0.84 and from 0.85 to 0.89 for Unet and diffusion respectively when comparing patient specific with generic training. Overall, the best performance was achieved using a patient specific reinforcement learning using diffusion MBIR. We also tested our methodologies (using patient specific training) for reconstruction of actual planar images obtained in the treatment of PSMA (Fig.3). SSIM values for the reconstructions using Unet and Diffusion with respect to the SPECT (half an hour later) were 0.65 and 0.73 respectively. We observed better organ and lesion delineation using diffusion MBIR when compared to Unet for real cases as shown in Fig.3.

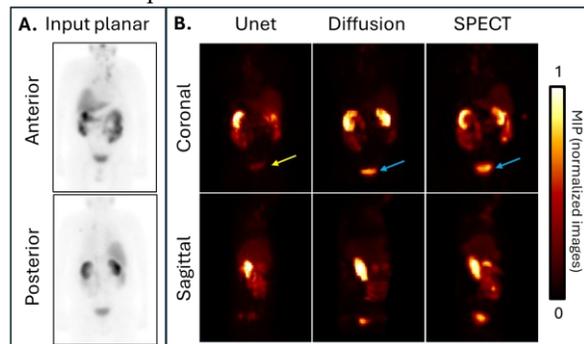

**Fig.3. A.** Acquired real planar acquisitions for $^{177}$Lu-PSMA scintigraphy. **B.** MIP at 0 (coronal) and 90 (sagittal) degrees after normalization of 3D reconstructions from the 2D planar images in (A) using patient specific reinforcement strategies: supervised learning (Unet) and unsupervised learning and diffusion MBIR (Diffusion). A SPECT acquisition was performed half an hour after the planar scintigraphy, the 3D SPECT is provided as baseline comparison. Arrows pointing to the prostate, showing better performance for diffusion MBIR than for the Unet approach.

### IV. DISCUSSION AND CONCLUSIONS

We have shown the feasibility of achieving 3D image reconstruction from 2D planar images using only anterior and posterior projections. We have explored different methodologies to achieve that reconstruction, finding best performance for diffusion MBIR with the generative model trained from a patient specific dataset. In terms of quantifications, we observed similar results for patient specific Unet and generic diffusion, this is due in part to the explicit enforcement in diffusion MBIR for the 3D solutions to represent the observed 2D planar measurements. However, we have observed (Fig. 2A) that not using patient specific training in diffusion makes the solution prone to generate unrealistic patient anatomies (e.g. with enlarged/reduced organs). On going work is focused on expanding the validation on more real cases, focusing on lesion identification, and on improving the models (e.g. including also the pre-treatment PET/CT explicitly during sampling). We are also evaluating our method for different levels of noise and bias (such as scatter) and exploring alternative networks (e.g. fully 3D diffusion). In summary, we believe that our methodology can set a change of paradigm for nuclear medicine dosimetry allowing for 3D quantification using only planar scintigraphy without the need of expensive and time-consuming SPECT leveraging the pre-therapy information of the patients.